\documentclass[
journal=jacsat, % 
manuscript=article]{achemso}

\usepackage{subfigure}
\usepackage{color}

\usepackage{url}
\usepackage{hyperref}

\author{J. M. de Sousa$^{1,3,*}$, T. Botari$^1$, E. Perim$^2$, R. A. Bizao$^{1,4}$, and Douglas S. Galvao$^{1,}$}

\affiliation{$^1$Instituto de F\'isica Gleb Wataghin, Universidade Estadual de Campinas, 13083-970, Campinas, SP, Brazil.}
\affiliation{$^2$Department of Mechanical Engineering and Materials Science, Duke University, Durham, North Carolina 27708, USA}
\affiliation{$^3$Departamento de F\'isica, Universidade Federal do Piau\'i, Teresina, Piau\'i, 64049-550, Brazil}
\affiliation{$^4$Department of Civil, Environmental and Mechanical Engineering, Laboratory of Bio-Inspired and Graphene Nanomechanics, University of Trento, via Mesiano, 77, 38123 Trento, Italy.}
\email{galvao@ifi.unicamp.br  ; josemoreiradesousa@gmail.com}

\title[\texttt{achemso} demonstration]
{Mechanical and Structural Properties of Graphene-like Carbon Nitride Sheets}

\begin{document}

\begin{abstract}

Carbon nitride-based nanostructures have attracted special attention (from theory and experiments) due to their remarkable electromechanical properties. In this work we have investigated the mechanical properties of some graphene-like carbon nitride membranes through fully atomistic reactive molecular dynamics simulations. We have analyzed three different structures of these $CN$ families, the so-called graphene-based g-$CN$, triazine-based g-$C_{3}N_{4}$ and heptazine-based g-$C_{3}N_{4}$. The stretching dynamics of these membranes was studied for deformations along their two main axes and at three different temperatures: $10K$, $300K$ and $600K$. We show that $g-CN$ membranes have the lowest ultimate fracture strain value, followed by heptazine-based and triazine-based ones, respectively. This behavior can be explained in terms of their differences in terms of density values, topologies and types of chemical bonds. The dependency of the fracture patterns on the stretching directions is also discussed.

\end{abstract}

\section{Introduction}

Due to the advent of nanotechnology, which created a new revolution in materials science, there is a renewed interest in organic and inorganic materials. Among these structures, carbon nitrides ($CN$) are of particular interest. Theoretical calculations have pointed out that CN crystals should present extremely high bulk modulus, of the order of $427$ GPa \cite{cohen1985calculation,liu1989prediction,liu1994stability}. For instance, $cubic-C_{3}N_{4}$ is predicted to exhibit higher bulk modulus than that of diamond\cite{teter1996low}, while $\beta-C_{3}N_{4}$ can exhibit a tunable electronic character, going from metallic to insulating depending on the morphology \cite{miyamoto1997theoretical}. These promising results motivated many different experimental investigations on distinct CN forms. Successful synthesis of materials like small $\beta-C_{3}N_{4}$ crystals, amorphous CN films \cite{niu1993experimental,yu1994observation} and nanofibers made from $C_{3}N_{4}$ and CN have been reported \cite{terrones1999carbon}.

 Due to the recent successfull isolation of graphene membranes \cite{novoselov2004electric}, with its unique electrical and mechanical properties\cite{berger2006electronic,neto2009electronic,geim2009graphene,lee2008measurement} and various applications in nanotechnology\cite{joshi2014precise,yang2013liquid,feng2013new,tao2013incorporating}, there has been a renewed interest in two dimensional materials. Other two dimensional structures, such as boron nitride\cite{auwarter1999xpd} and silicene\cite{vogt2012silicene}, among others, have been object of recent investigations. However, bidimensional $CN$ structures have not been thoroughly investigated, in spite of their very promising mechanical and electronic properties \cite{zheng2012graphitic,wang2009metal,wang2010excellent,perim2014novel,zheng2012graphitic}.

Questions about the actual synthesis of the carbon nitride graphitic phase still remain, however some evidence of its synthesis have been reported \cite{li2007synthesis,zhao2005turbostratic,wei2009synthesis,zelisko2014anomalous,algara2014triazine,thomas2008graphitic,Kroke2002ex}, while the synthesis of its polymeric phase (called melon) is well documented \cite{Lotsch2007,Doblinger2009}.
These materials are porous, low-density, hard, chemically inert, biocompatible structures\cite{cui2000review}, with unusual optical and electronic properties \cite{algara2014triazine,thomas2008graphitic,deifallah2008electronic,mo1999interesting}. These properties can be, in principle, exploited in a large class of technological applications.

\begin{figure}[htb!]
 \centering
 \includegraphics[width=0.60\linewidth]{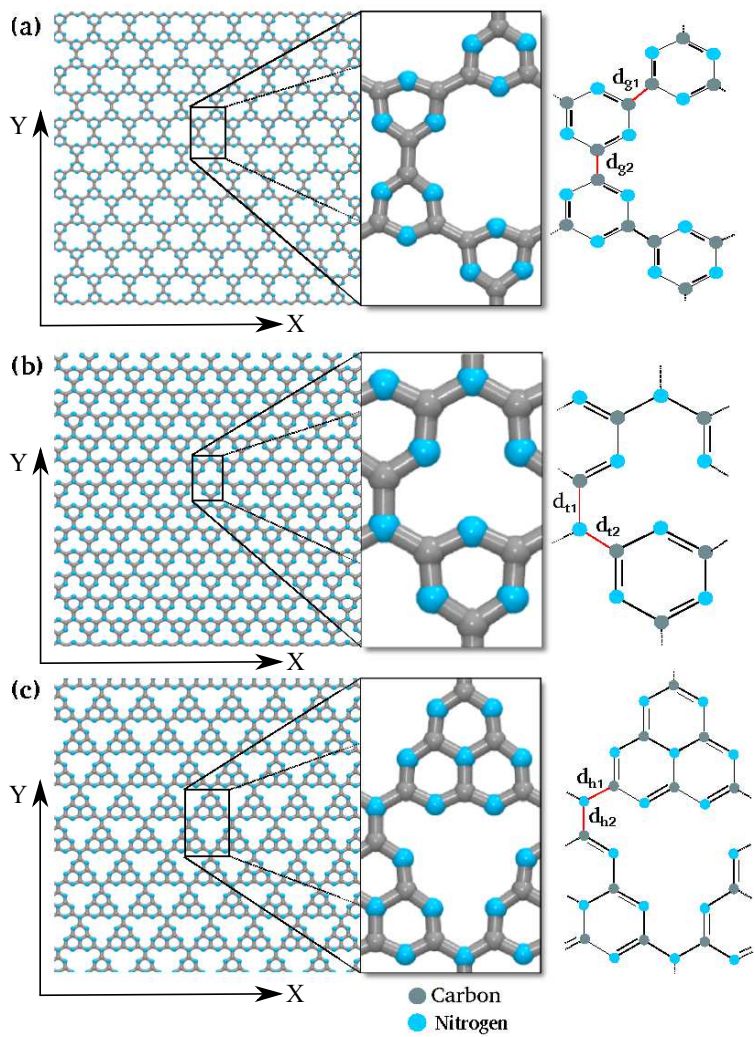}
 \caption{ Structural schemes of the investigated sheets: (a) graphene-based $g-CN$; (b) triazine-based $g-C_{3}N_{4}$, and; (c) heptazine-based $g-C_{3}N_{4}$ membranes. The insets show their corresponding unit cell and highlights some important bond-lengths. }
 
\label{Fig01}
\end{figure}

In this work we have investigated the mechanical and fracture patterns of three members of the two dimensional $CN$ family (Figure \ref{Fig01}): graphene-based g-$CN$, triazine-based g-$C_{3}N_{4}$ and heptazine-based g-$C_{3}N_{4}$. We have carried out fully atomistic reactive molecular dynamics simulations considering different stretching directions (along their main axes) and at different temperatures.

\section{Methodology}

All calculations were carried out with reactive classical molecular dynamics methods using the ReaxFF force field \cite{budzien2009reactive}, as implemented in the LAMMPS package \cite{plimpton1995fast}. ReaxFF was developed in order to simulate large systems while keeping an accurate description of bond formation and bond break processes. This method employs total energy description based on partial energy contributions, such as bond elongation, van der Waals forces and Coulomb interactions, among others. The ReaxFF ability to dynamically describe hybridization changes and charge redistribution (allowing the description of creating/breaking bonds), makes it suitable for the present study.

ReaxFF parameters are obtained from experiments and/or DFT calculations. The mean deviation between the heat of formation predicted by this method and experimental data is no larger than $2.9 kcal/mol$ for hydrocarbon systems\cite{van2001reaxff}. To further assess the suitability of the employed parameter set\cite{budzien2009reactive}, we compared the predicted structures of $\beta$-$C_{3}N_{4}$ and $graphitic$-$C_{3}N_{4}$ with other values previously reported in the literature\cite{teter1996low}. The diferences on bond-length and lattice parameter values were of $1$\% and $2$\% respectively, thus corroborating the adequacy of the used parameter set for this family of structures.

The investigated models consist of three different carbon nitride membranes called here: graphene-based g-CN, triazine-based g-C$_3$N$_4$ and heptazine-based g-C$_3$N$_4$, as shown in Figure \ref{Fig01}. The considered membranes in this work have dimensions around of $160$ $\times$ $150$ \AA, where the g-$CN$ structure has $6068$ atoms, triazine-based $g-C_{3}N_{4}$ and heptazine-based $g-C_{3}N_{4}$ ones, $9240$ and $8624$ atoms, respectively. All structures were considered with periodic boundary conditions along the $X$ and $Y$ directions. To assure that each structure was at equilibrium before the start of the stretching process, we first thermalized them. In order to do this, we ran the molecular dynamics simulations under the $NPT$ ensemble, i.e., with fixed number of atoms, pressure and temperature values. External pressure was set to zero, so we had no initial stress on any structure. The value of the chosen temperature was controlled during the stretching process through a Nos\'e-Hoover chain thermostat\cite{martyna1996explicit}. Three different temperatures values were considered ($10K$, $300K$ and $600K$), in order to determine how dependent the mechanical properties are on thermal effects.

The stretching process was simulated through the gradual increase of the lattice parameter along the periodic directions. A timestep of $0.05 fs$ was used together with a constant strain rate of $10^{-6}/fs$. The increased stretching is maintained until complete rupture of the membranes, which means tipical simulation times of the order of $10^{6}fs$. The methodology used in this work has been succeffully applied in the study of the mechanical properties of many other structures\cite{nair2011minimal,cranford2011mechanical,garcia2010bioinspired,Botari,jensen2015effect,de2016torsional}.

From the simulated stretching processes we can obtain the stress-strain curves.  In the linear region of the stress-strain curve we have calculated the Young's modulus, which can be defined as
\begin{equation}
Y = \frac{\sigma_{ii}}{\epsilon_i},
\end{equation}
where $\epsilon_i$ is the strain along direction $i$ and $\sigma_{ii}$ is the in-plane virial stress tensor component along direction $i$, defined as 
\begin{equation} 
\sigma_{ij}=\frac{\sum_{k}^{N}m_{k}v_{k_i}v_{k_j}}{V}+\frac{\sum_{k}^{N}r_{k_i}\cdot f_{k_j}}{V},
\end{equation}
where $V$ is the volume of the membrane, $N$ is the number of atoms, $v$ the velocity, $r$ the position and $f$ the force per atom. As the membrane is only one atom-thick and atomic volumes are not very well-defined, we opt to calculate all Young's moduli as a function of this thickness $d$, effectively writing the volume $V$ as $V=A.d$, where $A$ is the surface area of the membrane.

In order to have a better estimation of the spatial stress distribution during the stretching regime, we have  also calculated the von Mises stress per atom $i$, defined as
\begin{equation}
\resizebox{.85\hsize}{!}{
$\sigma_{vm}^i= \sqrt{\frac{\left(\sigma_{11}^i-\sigma_{22}^i\right)^2+\left(\sigma_{22}^i-\sigma_{33}^i\right)^2+\left(\sigma_{11}^i-\sigma_{33}^i\right)^2+6\left(\sigma_{12}^i{}^{2}+\sigma_{23}^i{}^{2}+\sigma_{31}^i{}^{2}\right)}{2}}$}.
\end{equation}

\section{Results and discussions}

 The three distinct membranes, graphene-based g$-CN$, triazine-based g$-C_3N_4$ and heptazine-based g$-C_3N_4$ (Figure \ref{Fig01}) were stretched at a constant rate until complete rupture. Table \ref{tab} summarizes the critical strain values (i.e., strain values at the point where fracture starts) for the three structures at three different temperatures, and for two distinct strain directions. We can see a clear difference for these values for each structure. This can be explained by the considerable difference in the strain energies associated with each stretched membrane. In the case of graphene-based g-$CN$, there are $C-C$ single bonds, which are naturally weaker than resonant $C-N$ bonds. The presence of these bonds decreases the strain energy associated with the stretched membrane, therefore making it easier to fracture. In the cases of heptazine-based and triazine-based g-$C_3N_4$, both present the same types of bonds, i.e., single and double $C-N$ bonds, but the pore density and, therefore, the number of these chemical bonds in their unit cells, is considerably different for each case. Heptazine-based structures show a lower density of chemical bonds than triazine-based structures, meaning a lower strain energy associated with the former. From these arguments, we can understand the variation on the critical strain values due to their different topologies. The decrease of the strain rate values with the increase of temperature is an expected effect, as higher thermal energy increases the fluctuations, thus making bond breaking easier.

\begin{figure}[htb!]
 \centering
 \includegraphics[width=0.60\linewidth]{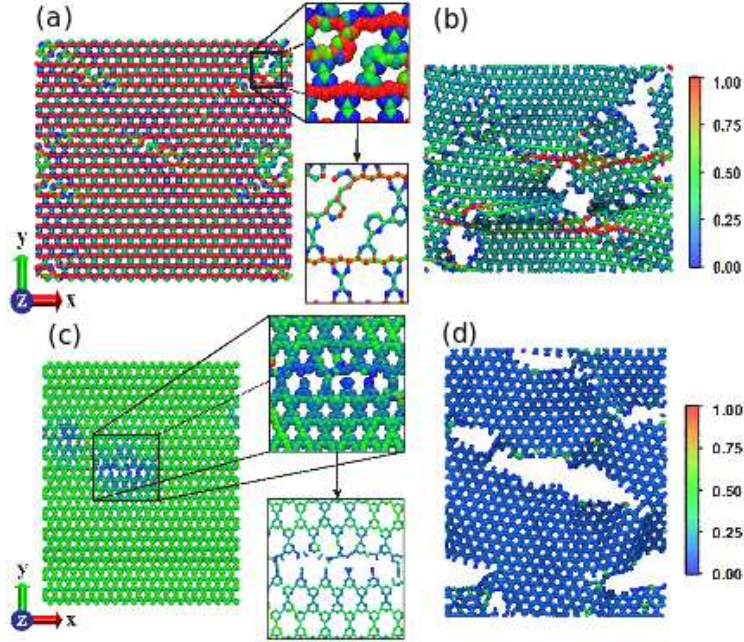}
 \caption{MD snapshots showing the stretch process considering (a-b) $X$ and (c-d) $Y$ directions for the g-$CN$ membrane. The insets show the beggining of the fracture process. The von Mises stress values indicate the stress distribution during the process by color scale labeled in the figure.}
\label{fg}
\end{figure}

\begin{figure}[htb!]
 \centering
 \includegraphics[width=0.60\linewidth]{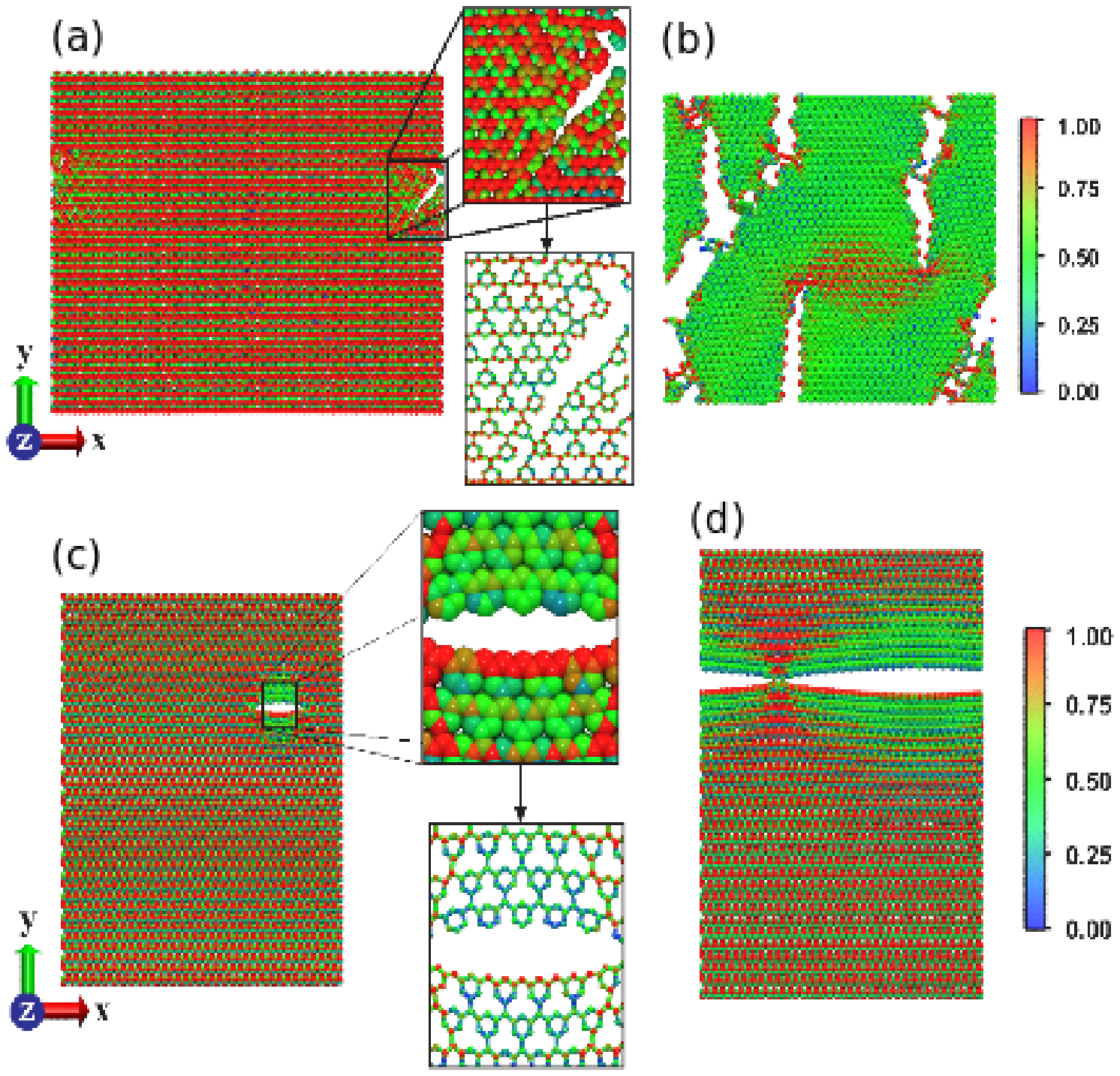}
 \caption{
 MD snapshots showing the stretch process considering (a-b) $X$ and (c-d) $Y$ directions for the triazine g-$C_{3}N_{4}$ membrane. The insets show the beggining of the fracture process. The von Mises stress values indicate the stress distribution during the process by color scale labeled in the figure.}
\label{ft}
\end{figure}

\begin{figure}[htb!]
 \centering
 \includegraphics[width=0.60\linewidth]{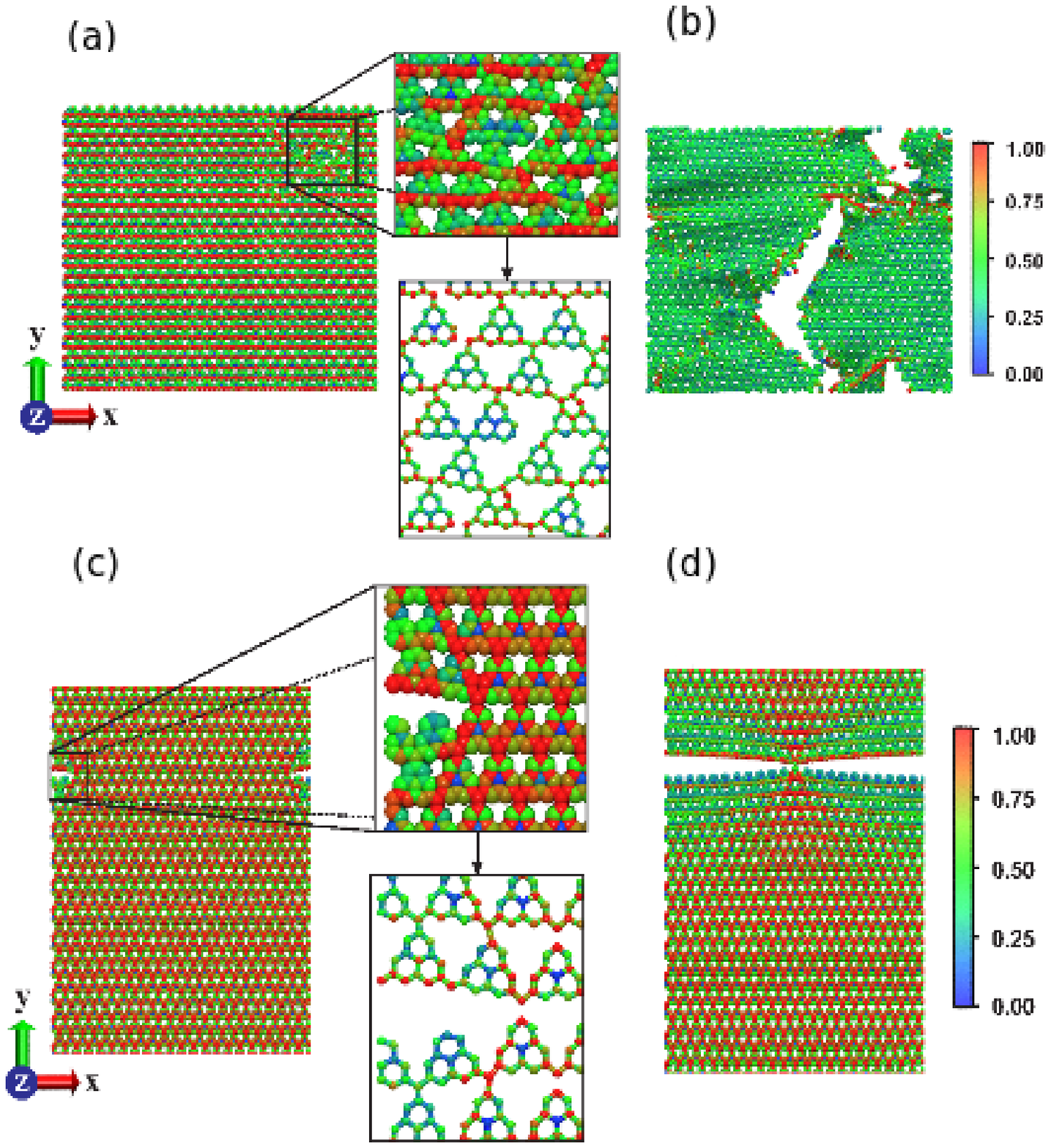}
 \caption{
  MD snapshots showing the stretch process considering (a-b) $X$ and (c-d) $Y$ directions for the heptazine g-$C_{3}N_{4}$ membrane. The insets show the beggining of the fracture process. The von Mises stress values indicate the stress distribution during the process by color scale labeled in the figure.}
 \label{fh}
\end{figure}

The distinct morphologies of each membrane type lead to different fracture patterns. Also, these patterns depend on the direction of the applied strain. We applied strain along the two principal directions, $X$ and $Y$, as defined in figures \ref{fg}, \ref{ft} and \ref{fh}. Figure \ref{fg} shows that, when stretching a graphene-based g-$CN$ sheet along the $X$ direction, stress acumulates on the $C-C$ single bonds that are almost parallel to that direction. These are the first bonds to break. When the strain is applied along the $Y$ direction, the C-C single bonds are parallel to the strecthing direction, thus much less stress is built up before the fracturing process starts, ultimately breaking these single bonds.

For triazine-based g-$C_3N_4$ membranes, as shown in figure \ref{ft}, the stress also builts up mostly into the single bonds, in this case $C-N$. When stretching along the $X$ direction, fracture yields rough edges, while stretching along the $Y$ direction fracture yields very clean edges. This is due to the single $C-N$ bonds which are aligned with that direction, breaking in sucession. 

A very similar behavior is observed in the case of heptazine-based g-$C_3N_4$ membranes, as shown in figure \ref{fh}. This should be expected as, despite presenting larger macro-cycles, the heptazine-based membranes present a very similar structure to that of triazine-based membranes. The types of chemical bonds are almost the same, as well as, their alignment with the stretching directions, X and Y. Therefore, while the critical strain values vary significantly between each of these structures, due to the different density (number) of chemical bonds, the stress and fracture patterns are very similar.

Stress strain curves for all the considered structures and temperatures are presented in Fig. 5. In general, the stress-strain curves start with a linear region where the Young's Modulus can be calculated, going through a non-linear region and until the total rupture. Analyzing the stress-strain curves, we observe that for the graphene-based g-CN membranes a direct transition from linear regime to the fracture occurs. For the triazine g-$C_{3}N_{4}$ and heptazine g-$C_{3}N_{4}$, after the linear region, the stress is momentarily relieved and another linear region can be observed, leading to a complete fracture afterwards.  This stress decrease can be attributed to an internal rearrangement of bond lengths and angles. Similar behavior in membranes formed by carbon, nitrogen and boron was observed using the Tersoff potential \cite{mortazavi2015mechanical}.

The Young's Modulus for all considered structures, directions and temperatures were obtained by fitting the linear region of the stress-strain curves. For instance, considering the room temperature ($300K$) and the $X$ direction, the obtained value for g-$CN$ was $1663$GPa.\AA, while for triazine g-$C_{3}N_{4}$ $1668$GPa.\AA and heptazine g-$C_{3}N_{4}$ $1247$GPa.\AA, as can be seen along with another results in Table \ref{tab2}. Our results are in good agreement with previously theoretical values obtained from a recent work with Tersoff potential for the case of triazine-based g-$C_{3}N_{4}$ \cite{mortazavi2015mechanical}.

Comparing the calculated value of the Young's modulus for graphene ($3570$GPa.\AA\cite{liu2007ab}) with the values herein reported, CN membranes values are lower by $53\%$ for g-$CN$ and triazine g-$C_{3}N_{4}$ and $65$\% for heptazine g-$C_{3}N_{4}$.
This decrease is due to differences in the chemical structure of the carbon nitride sheets when compared to graphene, namely the presence of pores, decreasing the density of chemical bonds, as well as the presence of single bonds.

\newpage
\newpage

\newpage

\begin{figure*}[top]
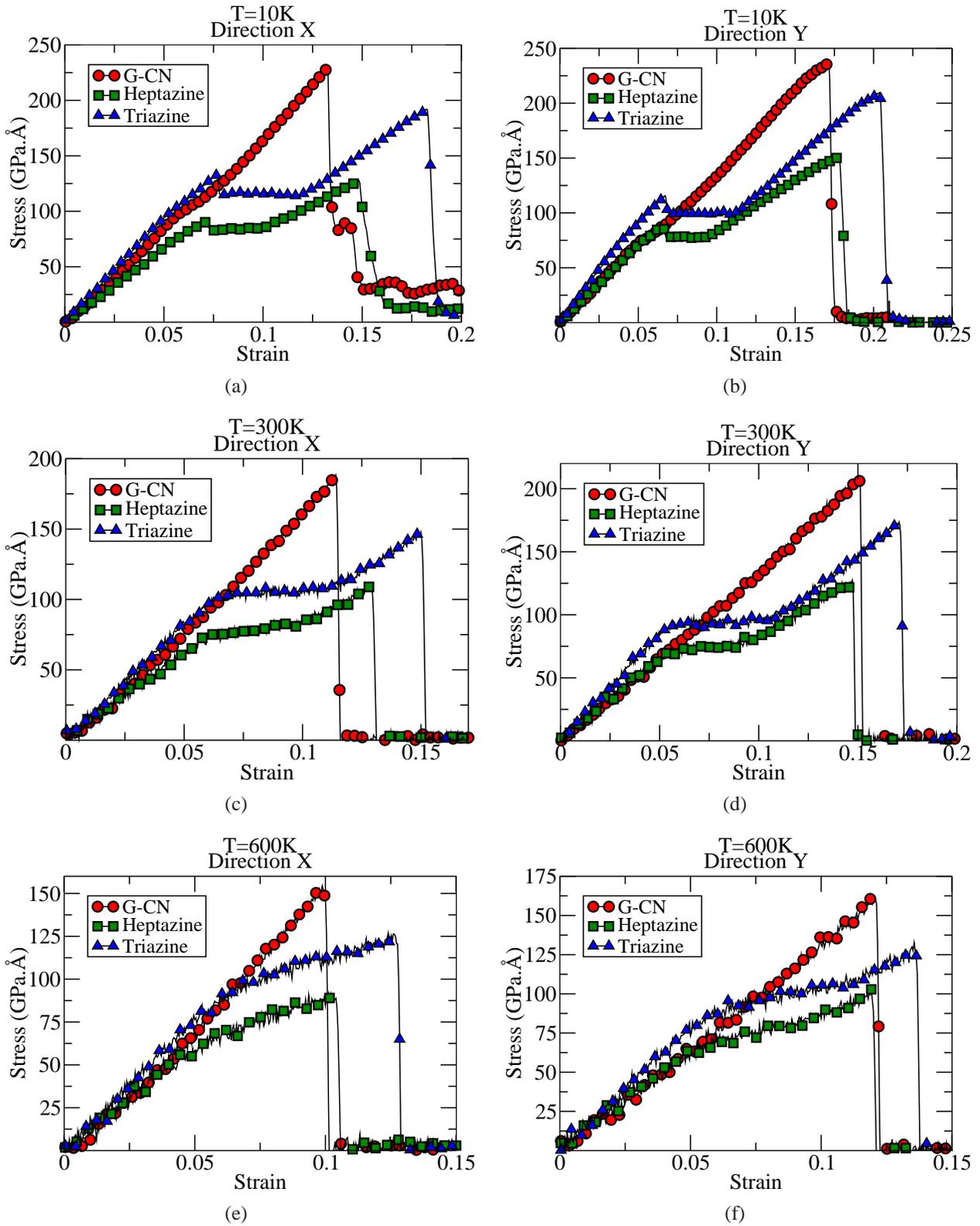

\centering
\mbox{\subfigure[]{\includegraphics[width=0.49\linewidth]{10k_x.eps}}\quad
\subfigure[]{\includegraphics[width=0.49\linewidth]{10k_y.eps} }}\\
\mbox{\subfigure[]{\includegraphics[width=0.49\linewidth]{300k_x.eps}}\quad
\subfigure[]{\includegraphics[width=0.49\linewidth]{300k_y.eps} }}\\
\mbox{\subfigure[]{\includegraphics[width=0.49\linewidth]{600k_x.eps}}\quad
\subfigure[]{\includegraphics[width=0.49\linewidth]{600k_y.eps} }}
\caption{Stress-strain curves of the carbon nitride sheets (g-$CN$, heptazine g-$C_{3}N_{4}$ and triazine g-$C_{3}N_{4}$) for different directions and temperatures.} 
\label{youngall}
\end{figure*}
\newpage

\newpage
\begin{table} 

\centering

\caption{Critical strain values.} 

\begin{tabular}{ccccc} 

\hline 

Direction & Temperature (K) & g-$CN$ & Heptazine g-$C_3N_4$ & Triazine g-$C_3N_4$\\ 

\hline
\hline

$X$ & 10 & 0.132 & 0.149 & 0.183\\
$Y$ & 10 & 0.172 & 0.178 & 0.204\\
\hline
\hline
$X$ & 300 & 0.114 & 0.129 & 0.150\\
$Y$ & 300 & 0.150 & 0.140 & 0.170\\
\hline
\hline
$X$ & 600 & 0.100 & 0.104 & 0.130\\
$Y$ & 600 & 0.120 & 0.120 & 0.137\\
\hline
\end{tabular}
\label{tab}
\end{table} 

\newpage

\begin{table} 

\centering

\caption{Young's modulus values.}

\begin{tabular}{cccc} 

\hline 

Structure & Temperature (K) & Young's Modulus (GPa.\AA) & Direction \\ 

\hline
\hline

G-CN & 10K & 1675 & X \\
Heptazine & 10K & 1356 & X \\
Triazine & 10K & 1890 & X \\
\hline
\hline
G-CN & 10K & 1516 & Y\\
Heptazine & 10K & 1397 & Y\\
Triazine & 10K & 1920 & Y\\
\hline
\hline
G-CN & 300K & 1663 & X\\
Heptazine & 300K & 1247 & X\\
Triazine & 300K & 1668 & X\\
\hline
\hline
G-CN & 300K & 1349 & Y\\
Heptazine & 300K & 1299 & Y\\
Triazine & 300K & 1733 & Y\\
\hline
\hline
G-CN & 600K & 1571 & X\\
Heptazine & 600K & 1197 & X\\
Triazine & 600K & 1578 & X\\
\hline
\hline
G-CN & 600K & 1333 & Y\\
Heptazine & 600K & 1229 & Y\\
Triazine & 600K & 1575 & Y\\

\hline
\end{tabular}
\label{tab2}
\end{table} 

\newpage
\newpage
\section{Summary and Conclusions}

We have investigated the mechanical and fracture patterns of a series of two-dimensional $CN$ structures: g-$CN$, triazine g-$C_{3}N_{4}$ and heptazine g-$C_{3}N_{4}$ (Figure \ref{Fig01}). The study was carried out through fully atomistic reactive molecular dynamics simulations using the ReaxFF force field. The Young's moduli for the carbon nitride membranes are smaller when compared with the Young's modulus for graphene. This can be understood by presence of pores and single C-N bonds in the carbon nitride membranes.
 More interestingly, graphene-based g-$CN$ goes abruptly from elastic to brittle behavior, while triazine and heptazine g-$C_3N_4$ structures go through significant structural reconstructions with multiple elastic stages. This differentiated behavior can be explained by the differences in the density of chemical bonds and how the rings are oriented in relation to the stretching directions, resembling an arch-type effect recently reported to silicene membranes \cite{Botari}.

\acknowledgement

This work was supported in part by the Brazilian Agencies CAPES, CNPq and FAPESP. The authors thank the Center for Computational Engineering and Sciences at Unicamp for financial support through the FAPESP/CEPID Grant $\sharp$ 2013/08293-7. J.M.S. acknowledges the support from CAPES through the Science Without Borders program (project number A085/2013).

\newpage
\bibliography{CNs}

\end{document}